%% file: main.tex
\documentclass[runningheads]{llncs}

\usepackage[T1]{fontenc}

\usepackage[table,xcdraw,dvipsnames]{xcolor}

\usepackage{times}

\usepackage{cite}
\usepackage{amsmath}
\usepackage{graphicx,calc}
\usepackage{xspace}
\usepackage{tikz}
    \usetikzlibrary{arrows}

\usepackage{booktabs}
\usepackage{colortbl}
\usepackage{xcolor}
\usepackage{tabularx}

\usepackage{subcaption}
\usepackage{hyperref}
\usepackage[hyphenbreaks]{breakurl}
\usepackage{wrapfig}
\usepackage{todonotes}

\begin{document}

\include{macros.tex}

\title{\seal: Symbolic Execution with Separation Logic}
\subtitle{(Competition Contribution)\vspace*{-8mm}}

\renewcommand{\orcidID}[1]{{\href{https://orcid.org/#1}{\protect\raisebox{3.25pt}{\protect\includegraphics{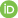}}}}}

\author{
    Tomáš Brablec
        \inst{1}
        \orcidID{0009-0005-4833-8662}
    \and
    Tomáš Dacík
        \thanks{Jury member.}
        \inst{1}
        \orcidID{0000-0003-4083-8943}
    \and 
    Tomáš Vojnar
        \inst{1,2}
        \orcidID{0000-0002-2746-8792}
}
\authorrunning{T. Brablec, T. Dacík and T. Vojnar}
\institute{ Faculty of Information Technology, Brno University of Technology,
Czech Republic \and Faculty of Informatics, Masaryk University, Czech
Republic\vspace*{-6mm}}

\maketitle

\begin{abstract} \seal is a static analyser for the verification of programs
that manipulate unbounded linked data structures. It is based on separation
logic to represent abstract memory states and, unlike other
separation-logic-based approaches, it employs a general-purpose separation logic
solver \astral for satisfiability and entailment checking, which itself is based
on translation to SMT. This design results in a modular architecture intended to
be easier to extend and to combine with reasoning in other theories. Although
still a prototype, \seal achieved competitive results in the
\textit{LinkedLists} base category and was one of only four analysers capable of
verifying programs with unbounded lists. We believe that the tool’s extensibility, combined with further development, can lead to significant
improvements in future competitions.\end{abstract}

\vspace*{-8mm}\section{Verification Approach}\vspace*{-1mm} \label{sec:overview} 

\enlargethispage{6mm}

\seal is a shape analyser based on forward abstract interpretation over formulae
of separation logic (SL) \cite{SL}, inspired by the approach of
\cite{SL_analysis}. Unlike the vast majority of SL-based analysers, \seal relies
on a general-purpose SL solver \astral \cite{astral} to check satisfiability and
entailment of SL formulae (instead of using some dedicated heuristic approach
for this purpose). As a result, analysis and logical reasoning in SL are cleanly
decoupled. The \astral solver was originally developed by the authors of this
paper and was further optimized and adapted for efficient handling of
verification queries within this work. Furthermore, since the decision procedure
implemented in \astral is based on a translation of SL into SMT, \seal can
seamlessly integrate SL reasoning with constraints from other theories.

As \seal is a relatively new tool, its potential has not yet been fully
exploited. Currently, our analyser supports only three built-in data structures
and simple integer properties. In the rest of the paper, we describe the
architecture of the tool and its current capabilities. More details can be found
in \cite{brablec_bp} (where a slightly earlier version of \seal is referred to
as \textsc{KTSN}).

\vspace*{-3mm}\paragraph{Abstract domain.}

\seal represents abstract states using so-called \emph{symbolic heaps}, i.e., SL
formulae of the form $\exists x_1, \ldots, x_m. \;\phi_1 * \ldots *
\phi_n$ where $*$ denotes a separating conjunction and each $\phi_i$ is an
atomic formula. Atomic formulae are either \emph{pure} describing
(dis)equalities of pointers or constraints over non-pointer values,
\emph{points-to} atoms describing a single heap cell, or \emph{inductive
predicates} used to describe unbounded sequences of pointers of a particular
shape.  A special predicate is used to explicitly represent memory that was
freed and is still reachable. The abstract domain can also represent precise
values of integer variables within the range $[-k, k]$ where $k = 5$ by default.
For each program location, a disjunction of symbolic heaps representing the
reachable memory configurations is computed and represented as a set of symbolic
heaps.

Currently, \seal supports three kinds of data structures: singly-linked lists
(SLLs), doubly-linked lists (DLLs), and nested singly-linked lists (NLLs). In
all three cases, the structure is acyclic and possibly empty. Some cyclic
structures can still be represented using an explicit representation of a
cyclic structure with multiple acyclic ones. 

\vspace*{-1mm}\paragraph{Abstraction.} 

\seal performs widening based on abstraction to generalise the shape of the memory by replacing its parts by instances of inductive predicates.
The abstraction is
performed by finding pairs of compatible spatial predicates (points-to or list
predicates) and joining them into a single list predicate---provided that the
memory location linking them is not the target of some external pointer links
and it is not pointed to by a program variable. This way of
abstraction is commonly used in other analysers, such as \predator
\cite{predator_SAS} or \cpachecker, too. The minimum length of a list predicate
is tracked up to a pre-defined, fixed limit. For example, the formula $\exists y.\;\mathsf{ls}_{1^+}(x,y) * y \mapsto z * \psi$ (where $\mathsf{ls}_{n^+}(x,y)$ denotes a list segment from $x$ to $y$ of length greater than or equal to $n$) is abstracted to $\mathsf{ls}_{2^+}(x,z) * \psi$ provided that (1)~$y$ does not occur in $\psi$ and (2) the original formula implies that $x \neq z$ (usually because $z$ is allocated in $\psi$) to guarantee acyclicity.

Currently, \seal guesses an expected inductive predicate for each C structure
before the analysis. A structure with one or two self-referential fields is
classified as an SLL or DLL, respectively. A structure with a single
self-referential field and an additional field previously classified
as an SLL is classified as an NLL. During the analysis, \seal just checks whether such
an abstraction can be performed.

The above approach can fail in cases like the following structure \texttt{TSLL}
defined as \texttt{\{struct TSLL* next; struct TSLL* inner;\}} and intended to
be used to build an NLL with the \texttt{next} pointer unused in the nested
lists. For this structure, \seal will guess its type to be DLL because it has
two self-referential fields. This weakness could be improved by selecting the
suitable inductive predicate for abstraction during the analysis. We would like
to improve this aspect in the future together with adding more predicates for
more data structures. We would also like to experiment with methods to
synthesize inductive predicates from examples.

\vspace*{-1mm}\paragraph{Finding loop fixpoints.}

The analysis reaches a fixpoint for a loop header location when the entailment
between the symbolic state $S$ from the current loop iteration and the state~$T$
from the previous loop iteration is valid. The entailment is
approximated by an incomplete method checking whether for each symbolic heap
$S_i \in S$ there exists $T_j \in T$ such that $S_i \models T_j$. The pairs
$(S_i, T_j)$ to be checked are chosen according to a heuristic based on the
numbers of allocated memory locations in $S_i$ and $T_j$. For an entailment to
be valid, there cannot be more allocations on the right side of the entailment
than there are on the left side. The same method is used for pruning redundant
symbolic heaps in the join operator. Additionally, a number of syntactic
simplifications are applied to the states during analysis, such as reduction of
equivalence classes in formulae, removal of variables outside of their scope, syntactic deduplication of formulae, etc.

\enlargethispage{6mm}

\vspace*{-1mm}\paragraph{Integer variables.}

Currently, \seal can only track concretely known values of integer variables
that are within a pre-configured range. Conditions and arithmetic operations on
these variables can be evaluated precisely. When the result of an operation on
such values goes out of the represented range, the variable that should receive
the result becomes unconstrained instead. The range of tracked integer values
can be configured, increasing the representable range trades precision for
length of analysis.

\vspace*{-1mm}\paragraph{Interprocedural analysis.}

Analysis of function calls is optimised by storing function summaries -- pairs
of formulae representing precondition and postcondition of the analysed
function. Using this information, a consecutive call of a function may be
replaced by applying a matching summary.

\vspace*{-1mm}\paragraph{Analysis results.}

\seal is able to verify absence of NULL-pointer dereferences and use-after-free
errors (\texttt{valid-deref} property), double-free errors (\texttt{valid-free}
property), and some memory leaks (those that fall under the
\texttt{valid-memtrack} property). The \texttt{valid-memcleanup} property is not
yet supported. When an issue is found after reaching a condition that could not
be precisely evaluated based on the current state 
(e.g., because of missing integer information), the analyser
returns the \texttt{UNKNOWN} result. However, false positives are still possible
because of the abstraction. In the future, they could be
avoided by validating the issue by rerunning
the analyser without abstraction. When an unsupported language feature 
is reached, the \texttt{ERROR} result is returned.

\vspace*{-1mm}\paragraph{SL solver.}

We briefly describe how the backend solver \astral of \seal handles SL queries.
Given an SL formula, $\astral$ first computes a bound on how many times it
suffices to unfold the predicate instances present in the given SL formula and
then uses this bound to translate the SL formula into several formulae in SMT
theories of bitvectors and arrays. The translated formulae are then passed to an
SMT solver, which is configured to be \bitwuzla \cite{bitwuzla} for SV-COMP.
Unlike in \cite{astral}, \astral runs in a mode supporting user-defined
inductive predicates, i.e., the predicates are defined by \seal and \astral runs
without any built-in support for them \cite{astral_cav}. 

\enlargethispage{6mm}

\vspace*{-2mm}\section{Strengths and Weaknesses}\vspace*{-1mm}
\label{sec:strenghts-and-weakneses}

\begin{table*}[!t]
  \centering

  \caption{The results of selected tools for \textit{MemSafety-LinkedLists} and
  its subset with unbounded lists. The columns give numbers of correctly
  verified programs and found bugs, numbers of programs with no result reported
  (error, unknown result or out of resources), and the total time used for
  programs for which a result was produced.}
    
  \label{table:results} 
  \input{tables/results.tex} 
\end{table*}

Table~\ref{table:results} provides a view on the results of
\textit{MemSafety-LinkedLists} \cite{SVCOMP26}, the only base category in which \seal
participates, illustrating the strengths of \seal. Since almost half of the
programs in this category manipulate lists of bounded size only and can
therefore be verified using bounded model checking (or similar heuristics), we
focus on the subset of benchmarks that require reasoning about unbounded data
structures. Specifically, we exclude the benchmark sets \texttt{list-simple} and
\texttt{list-ext3-properties}, which operate exclusively on bounded lists, as
well as \texttt{ddv-machzwd} where the list-manipulating code is unreachable.
Only the tools that can verify correctness of at least one of the remaining
programs are shown in the table.

%

\vspace*{-1mm}\paragraph{Strengths.}

As Table \ref{table:results} shows, out of 20 tools (excluding metaverifiers)
participating in \textit{MemSafety}, only 4 of them were able to verify correctness
of programs with unbounded data structures. This indicates that reasoning about
such programs remains a challenging task. \seal can verify more programs than
\twols \cite{2ls_fmcad, 2LS-SVCOMP23}, but less than \predator
\cite{predator_SAS, PREDATORHP-SVCOMP20} and \cpachecker \cite{CPAchecker,
CPACHECKER-SVCOMP24}, two analysers based on \emph{symbolic memory graphs}
(SMG). We believe that the biggest strength of \seal lies in its potential to be
extended for data structures that cannot be represented by SMGs, such as
skip-lists or trees. Indeed, \astral in its new versions supports user-defined inductive predicates, including various kinds of trees, which may be implemented in \seal or potentially learned automatically. We would also like to use the generality of our technique
to create a witness validator for correctness memory safety witnesses, once a
witness format is established.

\vspace*{-1mm}\paragraph{Weaknesses.}

As an early prototype, \seal comes with a handful of limitations. First of all,
we support only a subset of the C language (most kinds of pointer arithmetic,
integer fields in list structures, global and static variables, arrays, and
other language features are not supported). For this reason, we do not
participate in other base categories of \textit{MemSafety}. \seal currently has
limited support for non-pointer values, only tracking precise values of integer
variables, where known. For this reason, we are often not able to report a
memory safety violation because an unknown condition in the program was reached,
making any violation result a possible false positive. This limitation can be
eliminated by tracking more complex information in symbolic states or by
embedding our analyser as a new abstract domain into the Frama-C's abstract
interpreter EVA~\cite{eva}.

\vspace*{-2mm}\section{Software Project and Tool Setup}\vspace*{-2mm}

\enlargethispage{6mm}

\seal is implemented in the OCaml language as a plugin of the Frama-C platform
\cite{Frama-C1, Frama-C2} (version \texttt{31.0}). Currently, it is a
stand-alone plugin that does not communicate with other analysers of Frama-C.
\seal relies on the Frama-C's intermediate representation (a fork of the CIL
project \cite{cil}) and data flow analysis engine. Optionally, \seal can
visualise its detailed results using the Frama-C's new GUI frontend Ivette
\cite{ivette}. We used the \astral solver version \texttt{2.0-prerelease.1} that
internally uses the \bitwuzla SMT solver version \texttt{0.8.2} through its
OCaml binding.

\vspace*{-1mm}\paragraph{Usage.}

A binary of \seal is available at Zenodo \cite{seal-zenodo} (the only runtime
dependency is \textsc{GCC}). In SV-COMP, \seal is run via a Python3 script
\texttt{seal-entrypoint.py}:
\begin{center}
  \texttt{./seal-entrypoint.py -machdep=[gcc\_x86\_32|gcc\_x86\_64] input.c}
\end{center}
\noindent where the machine models \texttt{gcc\_x86\_32} and \texttt{gcc\_x86\_64}
correspond to the architectures ILP32 and LP64 used by SV-COMP, respectively. See
the attached file \texttt{README.md} for more information.


\vspace*{-1mm}\paragraph{Software project.}

\seal is available under the MIT license, the project is maintained by Tomáš Brablec. The \astral solver which was modified as a part of this work is maintained by Tomáš Dacík.

\vspace*{-1mm}\paragraph{Participation.}

\seal participated in the \textit{MemSafety-LinkedLists} base category only.



\vspace*{-1mm}\paragraph{Data-Availability Statement.}

\seal is available under the MIT license at
\url{https://github.com/pepega007xd/seal}. The version participating in
SV-COMP'26 is available under the tag \texttt{svcomp26} and archived at Zenodo
\cite{seal-zenodo}.


\vspace*{-1mm}\paragraph{Acknowledgements.}

\newlength\myheight
\newlength\mydepth
\settototalheight\myheight{Xygp}
\settodepth\mydepth{Xygp}
\setlength\fboxsep{0pt}
\newcommand*\inlinegraphics[1]{%
  \settototalheight\myheight{Xygp}%
  \settodepth\mydepth{Xygp}%
  \raisebox{-\mydepth}{\includegraphics[height=\myheight]{#1}}%
}

We appreciate discussion with members of the Frama-C team at CEA, France. The research was supported by the Czech Science Foundation project \mbox{26-22640S}. Tomáš Dacík was supported by the Brno PhD Talent Scholarship of the Brno City Municipality. The collaboration with CEA was supported by the project \mbox{VASSAL:} ``Verification and Analysis for Safety and Security of Applications in Life'' funded by the European Union under Horizon Europe WIDERA Coordination and
support Action/Grant Agreement No. 101160022. \raisebox{0.5mm}{\inlinegraphics{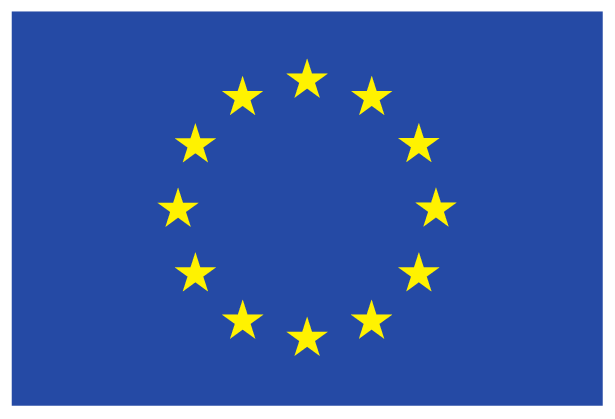}}

\bibliographystyle{splncs04}
\bibliography{references.bib}
\end{document}

%% file: macros.tex
\renewcommand{\phi}{\varphi}

\newcommand{\seal}{\textsc{Seal}\xspace}
\newcommand{\cpachecker}{\textsc{CPAchecker}\xspace}
\newcommand{\twols}{\textsc{2ls}\xspace}
\newcommand{\predator}{\textsc{PredatorHP}\xspace}

\newcommand{\astral}{\textsc{Astral}\xspace}
\newcommand{\bitwuzla}{\textsc{Bitwuzla}\xspace}

%% file: tables/results.tex
\setlength{\tabcolsep}{5.5pt} 

\begin{tabularx}{\textwidth}{
  l
  *{4}{r}
  c
  *{4}{r}
}
\toprule
& \multicolumn{4}{c}{All programs (134)} & &
  \multicolumn{4}{c}{Unbounded (69)}\\
\cmidrule(lr){2-5} \cmidrule(lr){7-10}

Verifier & True & False & No res. & Time [s] & & True & False & No res. & Time [s]\\
\midrule

\rowcolor[gray]{0.9}
\predator & 96 & 28 & 10 & 112 && 46 & 14 & 9 & 77\\
\cpachecker & 87 & 27 & 20 & 2158 && 40 & 13 & 16 & 1550\\
\rowcolor{GreenYellow}
\seal & 55 & 4 & 75 & 571 && 14 & 2 & 53 & 166\\
\twols & 46 & 14 & 73 & 1154 && 3 & 13 & 53 & 92\\

\bottomrule
\end{tabularx}

%% file: main.bbl
\begin{thebibliography}{10}
\providecommand{\url}[1]{\texttt{#1}}
\providecommand{\urlprefix}{URL }
\providecommand{\doi}[1]{https://doi.org/#1}

\bibitem{CPACHECKER-SVCOMP24}
Baier, D., Beyer, D., Chien, P.C., Jankola, M., Kettl, M., Lee, N.Z., Lemberger, T., Lingsch-Rosenfeld, M., Spiessl, M., Wachowitz, H., Wendler, P.: {CPAchecker 2.3 with Strategy Selection (Competition Contribution)}. In: Proc. of TACAS~(3). pp. 359--364. LNCS~14572, Springer (2024). \doi{10.1007/978-3-031-57256-2_21}

\bibitem{CPAchecker}
Baier, D., Beyer, D., Chien, P.C., Jakobs, M.C., Jankola, M., Kettl, M., Lee, N.Z., Lemberger, T., Lingsch-Rosenfeld, M., Wachowitz, H., Wendler, P.: {Software Verification with CPAchecker 3.0: Tutorial and User Guide}. In: Proc. of FM. pp. 543--570. LNCS~14934, Springer (2024). \doi{10.1007/978-3-031-71177-0_30}

\bibitem{Frama-C2}
Baudin, P., Bobot, F., B\"{u}hler, D., Correnson, L., Kirchner, F., Kosmatov, N., Maroneze, A., Perrelle, V., Prevosto, V., Signoles, J., Williams, N.: {The Dogged Pursuit of Bug-Free C Programs: The Frama-C Software Analysis Platform}. Commun. ACM  \textbf{64}(8),  56--68 (2021). \doi{10.1145/3470569}

\bibitem{SVCOMP26}
Beyer, D., Strejček, J.: {Evaluating Software Verifiers for {C}, {Java}, and {SV-LIB} (Report on {SV-COMP 2026})}. In: Proc. of TACAS~(2). LNCS~16506, Springer (2026)

\bibitem{eva}
Blazy, S., B{\"u}hler, D., Yakobowski, B.: {Structuring Abstract Interpreters Through State and Value Abstractions}. In: Proc. of VMCAI. pp. 112--130. LNTCS~10145, Springer (2017). \doi{10.1007/978-3-319-52234-0_7}

\bibitem{brablec_bp}
Brablec, T.: Static Analysis of Heap-Manipulating Programs using Separation Logic. Bachelor's thesis, Faculty of Information Technology, Brno University of Technology, Brno (2025), \url{http://hdl.handle.net/11012/253189}

\bibitem{seal-zenodo}
Brablec, T., Dacík, T., Vojnar, T.: {SEAL (SV-COMP 2026)}. Zenodo (2025). \doi{10.5281/zenodo.17690232}

\bibitem{ivette}
Correnson, L.: {Ivette: A Modern GUI for Frama-C}. In: Proc. of SEFM. pp. 116--131. LNCS~13765, Springer (2022). \doi{10.1007/978-3-031-26236-4_10}

\bibitem{Frama-C1}
Cuoq, P., Kirchner, F., Kosmatov, N., Prevosto, V., Signoles, J., Yakobowski, B.: {Frama-C: A~Software Analysis Perspective}. In: Proc. of SEFM. pp. 233--247. LNPSE~7504, Springer (2012). \doi{10.1007/978-3-642-33826-7_16}

\bibitem{astral}
Dac{\'i}k, T., Rogalewicz, A., Vojnar, T., Zuleger, F.: {Deciding Boolean Separation Logic via Small Models}. In: Proc. of TACAS. pp. 188--206. LNCS~14570, Springer (2024). \doi{10.1007/978-3-031-57246-3_11}

\bibitem{astral_cav}
Dac\'{\i}k, T., Rogalewicz, A., Sighireanu, M., Vojnar, T., Zuleger, F.: {Deciding Separation Logic with Inductive Definitions via Small Models and Translation to SMT} (2026), under preparation for submission.

\bibitem{SL_analysis}
Distefano, D., O'Hearn, P.W., Yang, H.: {A Local Shape Analysis Based on Separation Logic}. In: Proc. of TACAS. pp. 287--302. LNCS~3920, Springer (2006). \doi{10.1007/11691372_19}

\bibitem{predator_SAS}
Dudka, K., Peringer, P., Vojnar, T.: {Byte-Precise Verification of Low-Level List Manipulation}. In: Proc. of SAS. pp. 215--237. LNPSE~7935, Springer (2013). \doi{10.1007/978-3-642-38856-9_13}

\bibitem{2ls_fmcad}
Mal\'{\i}k, V., Hru\v{s}ka, M., Schrammel, P., Vojnar, T.: {Template-Based Verification of Heap-Manipulating Programs}. In: Proc. of FMCAD. p. 103–111. IEEE (2018). \doi{10.23919/FMCAD.2018.8603009}

\bibitem{2LS-SVCOMP23}
Malík, V., Schrammel, P., Vojnar, T., Nečas, F.: {2LS: {Arrays} and Loop Unwinding (Competition Contribution)}. In: Proc. of TACAS~(2). pp. 529--534. LNCS~13994, Springer (2023). \doi{10.1007/978-3-031-30820-8_31}

\bibitem{cil}
Necula, G.C., McPeak, S., Rahul, S.P., Weimer, W.: {CIL: Intermediate Language and Tools for Analysis and Transformation of C Programs}. In: Proc. of CC. pp. 213--228. LNCS~2304, Springer (2002). \doi{10.1007/3-540-45937-5_16}

\bibitem{bitwuzla}
Niemetz, A., Preiner, M.: Bitwuzla. In: Proc. of {CAV}. pp. 3--17. LNCS~13965, Springer (2023). \doi{10.1007/978-3-031-37703-7\_1}

\bibitem{PREDATORHP-SVCOMP20}
Peringer, P., Šoková, V., Vojnar, T.: {PredatorHP Revamped (Not Only) for Interval-Sized Memory Regions and Memory Reallocation (Competition Contribution)}. In: Proc. of TACAS~(2). pp. 408--412. LNCS~12079, Springer (2020). \doi{10.1007/978-3-030-45237-7_30}

\bibitem{SL}
Reynolds, J.: {Separation Logic: A Logic for Shared Mutable Data Structures}. In: Proc. of LICS. pp. 55--74. IEEE (2002). \doi{10.1109/LICS.2002.1029817}

\end{thebibliography}
